\documentclass[prb, superscriptaddress]{revtex4}
\usepackage{graphicx}
\usepackage{color}
\usepackage{amsmath}
\usepackage{amsfonts}
\usepackage{enumitem}
\usepackage{hyperref}
\hypersetup{
	colorlinks = true,
	citecolor = blue,
	linkcolor= blue
}

\newcommand{\be}{\begin{equation}}
\newcommand{\ee}{\end{equation}}

\newcommand{\bea}{\begin{eqnarray}}
\newcommand{\eea}{\end{eqnarray}}

\newcommand{\p}{\partial}

\newcommand{\la}{\left\langle}

\renewcommand{\vec}[1]{{\boldsymbol #1}}

\renewcommand{\phi}{\varphi}
\renewcommand{\epsilon}{\varepsilon}
\def\nn{\nonumber\\}

\usepackage[normalem]{ulem}

\begin{document}

\title{Electronic states of pseudospin-1 fermions in $\alpha-\mathcal{T}_3$ lattice ribbons in a magnetic field}
\date{\today}

\author{O. V. Bugaiko}
\affiliation{Department of Physics, Taras Shevchenko National University of Kiev, Kiev, 03680, Ukraine}

\author{D. O. Oriekhov}
\affiliation{Department of Physics, Taras Shevchenko National University of Kiev, Kiev, 03680, Ukraine}

\begin{abstract}
	The electronic states on a finite width $\alpha-\mathcal{T}_3$ ribbon in
	a magnetic field are studied in the framework of low-energy effective theory. Both zigzag and armchair types of boundary conditions are
	analyzed. The analytical solutions are compared with the results of numerical tight-binding calculations. It is found that the
	flat band of zero energy survives for all types of boundary conditions. The analytical estimates for the spectral gap in a weak 
	magnetic field are discussed. For zigzag type boundary conditions the approximate expressions for the edge and bulk electron states in the strong magnetic field are found.
\end{abstract}
%\pacs{81.05.ue, 73.22.Pr}
\maketitle

\section{Introduction}

After the experimental discovery of graphene [\onlinecite{Geim_graphene}] the systems with relativisticlike quasiparticle spectrum
attracted a great interest. In addition, it was shown [\onlinecite{Bradlyn}] that in crystals with special space groups one can
obtain fermionic excitations with no analogues in high-energy physics. One of the remarkable features of such quasiparticles is a possibility to
possess strictly flat bands [\onlinecite{Heikkila}] (for a recent review of artificial flat band systems, see Ref.[\onlinecite{Leykam}]). The 
dice model is probably historically the first example of such a system  with a flat band which hosts pseudospin-1 fermions 
[\onlinecite{Sutherland}].

Recently the $\alpha-\mathcal{T}_3$ model attracted a significant interest as an interpolation between graphene 
and dice model  [\onlinecite{Raoux}].
The $\alpha-\mathcal{T}_3$ model is a tight-binding model of two-dimensional fermions living on the $\mathcal{T}_3$ (or dice)
lattice where atoms are situated at both the vertices of a hexagonal lattice and the hexagons centers [\onlinecite{Sutherland,Vidal}].  The
parameter $\alpha$ describes the relative strength of the coupling between the honeycomb lattice sites and the central site. Since
the $\alpha-\mathcal{T}_3$ model has three sites per unit cell, the electron states in this model are described by three-component
pseudospin-1 fermions. It is natural then that the spectrum of the model is comprised of three bands.
 The two of them form a Dirac cone as in graphene, and the third band is completely flat and has zero energy [\onlinecite{Raoux}].
 All three bands meet at the $K$ and $K^{\prime}$ points, which are situated at the corners of the Brillouin zone.
 The $\mathcal{T}_3$ lattice has been experimentally realized in Josephson arrays [\onlinecite{Serret}] and its optical realization by laser beams was proposed in Ref.[\onlinecite{Rizzi}].   Recently, a 2D model for $\text{Hg}_{1-x}\text{Cd}_{x}\text{Te}$ at critical doping has been shown to map onto the $\alpha-\mathcal{T}_3$
model with an intermediate parameter $\alpha=1/\sqrt{3}$ [\onlinecite{Malcolm}].

The presence of completely flat energy band results in surprisingly strong paramagnetic response in a magnetic field in the dice model ($\alpha=1$) [\onlinecite{Raoux}]. The minimal conductivity and topological Berry winding were analyzed in three-band semi-metals in Ref. [\onlinecite{Louvet}]. The dynamic polarizability of the dice model was calculated in the random phase approximation [\onlinecite{Malcolm_2016}] and it was found that the plasmon branch due to strong screening in the flat band is pinched to the point $\omega=|\vec{k}|=\mu$. In addition, the singular nature of the Lindhard function leads to a much faster decay of the Friedel oscillations. Recently several physical quantities have been studied in the
	$\alpha-{\cal T}_3$ lattice such as orbital susceptibility [\onlinecite{Raoux}], optical and magneto-optical conductivity [\onlinecite{Malcolm_2014, Carbotte,Illes,Cserti}], magnetotransport [\onlinecite{Malcolm,Biswas,Xu,Islam}]. The role of transverse magnetic field on \textit{zitterbewegung} was studied in Ref.[\onlinecite{Biswas2018}]. The enhancement of thermoelectric properties of $\alpha-\mathcal{T}_3$ model was discussed in a recent paper [\onlinecite{Firoz_Islam}].

Perfectly flat bands are expected to be not stable with respect to generic perturbations. There are several most common types of such perturbations: the presence of boundaries, magnetic field, Coulomb impurities, and disorder. In a recent paper [\onlinecite{Oriekhov}], we showed that, remarkably, the energy dispersion of the completely flat
energy band of the dice model is not affected by the presence of boundaries except the trivial reduction of the number of
degenerated electron states due to the finite spatial size of the system. It was shown also that the
flat band for the dice model remains unaltered in the presence of circularly polarized radiation [\onlinecite{Dey}] and magnetic field [\onlinecite{Bercioux}]. Recently it was shown by numerical tight-binding calculations [\onlinecite{Chen_nanoribbons}] that the flat band in
the $\alpha-\mathcal{T}_3$ model survives even in the presence of a magnetic field and boundaries. Additionally, the formation of chiral edge states was predicted [\onlinecite{Dey_2019}] for nanoribbons made of $\alpha-\mathcal{T}_3$ lattice in the presente of circularly polarized light.  In this paper we determine exact analytical solutions for the low-energy electron states
of $\alpha-\mathcal{T}_3$ ribbons with different types of terminations in a magnetic field. This allows us to find analytical estimations for energy gaps opened by weak magnetic field. In the strong magnetic field limit (wide strip limit) we find the approximate expressions for the edge and bulk electron states spectrum.

The paper is organized as follows. The $\alpha-\mathcal{T}_3$ model and its electron states in infinite system are described in Sec.\ref{sec:model}. The electron states and energy spectra in ribbons with zigzag and armchair edges in a magnetic field are studied in Secs.\ref{sec:zigzag} and \ref{sec:armchair}, respectively. In Appendices \ref{appendix:add_zigzag} and \ref{appendix:armchair_add} we add extended discussion and give detailed derivations for previously mentioned cases. The analytical predictions are compared with the results of numerical tight-binding calculations for each type of termination. The summary of our results is given in Sec.\ref{sec:summary}.

\section{General properties of the model in external magnetic field}
\label{sec:model}

\begin{figure}
	\includegraphics[scale=0.4]{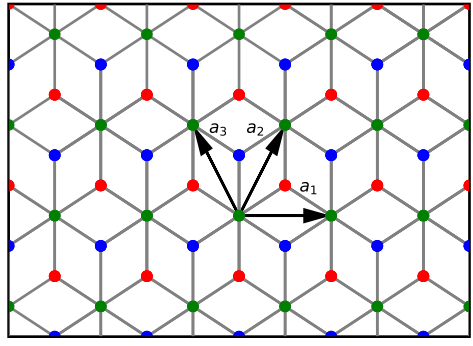}
	\caption{The ${\cal T}_3$ lattice whose red points display the atoms of the
		$A$ sublattice, the blue points describe the $B$ sublattice, and the green points define the $C$ sublattice. The vectors
		$\vec{a}_1=(\sqrt{3},\,0)a$ and $\vec{a}_2=(\sqrt{3}/2,\,3/2)a$ are the basis vectors of the $C$ sublattice.}
	\label{fig1}
\end{figure}

The $\alpha-\mathcal{T}_3$ model describes quasiparticles in two dimensions with pseudospin $S=1$ on the $\mathcal{T}_3$ lattice schematically shown in
Fig.\ref{fig1} [\onlinecite{Raoux}]. This lattice has a unit cell with three different lattice sites whose two sites ($A,C$) like in graphene form a honeycomb
lattice with hopping amplitude $t_{AC}=t_1$ and additional $B$ sites at the center of each hexagon are connected to the $C$ sites with
hopping amplitude $t_{BC}=t_2$. The two hopping parameters $t_1$ and $t_2$ are not equal, in general, and the dice model corresponds to the limit $t_1=t_2$. The lattice structure and basis vectors are shown on Fig.\ref{fig1}.

\subsection{Hamiltonian of the model}
\label{sec:Hamiltonian}

The corresponding tight-binding Hamiltonian in momentum space reads [\onlinecite{Raoux}]
\begin{align}
\label{TB-Hamiltonian}
H=\left(\begin{array}{ccc}
0 & f_{\vec{k}}\cos\Theta & 0\\
f^{*}_{\vec{k}}\cos\Theta & 0 & f_{\vec{k}}\sin\Theta\\
0 & f^{*}_{\vec{k}}\sin\Theta & 0
\end{array}\right),
\quad \alpha \equiv \tan\Theta=\frac{t_2}{t_1},\quad f_{\vec{k}}=-\sqrt{t_1^2+t_2^2}\,(1+e^{-i\vec{k}\vec{a}_2}+e^{-i\vec{k}\vec{a}_{3}}).
\end{align}
It is easy to find the energy spectrum of the above Hamiltonian, which is qualitatively the same for any $\alpha$ and consists of
three bands: the zero-energy flat band, $\epsilon_0(\mathbf{k})=0$, and two dispersive bands
\begin{equation}
\epsilon_{\pm}(\mathbf{k})=\pm|f_{k}|=\pm \sqrt{t_1^2+t_2^2}\bigg[3+2(\cos(\vec{a}_1\vec{k})+\cos(\vec{a}_2\vec{k})+
\cos(\vec{a}_3\vec{k}))\bigg]^{1/2}.
\end{equation}
The presence of a completely flat band with zero energy is perhaps one of the remarkable properties of the $\alpha-\mathcal{T}_{3}$ lattice
model.

There are six values of momentum for which $f_{\vec{k}}=0$ and all three bands meet. They are situated at the corners of the hexagonal
Brillouin zone. The two inequivalent points, for example, are
\begin{align}
\vec{K}=\frac{2\pi}{a}\left(\frac{\sqrt{3}}{9},\,\frac{1}{3}\right),\quad \vec{K}'=\frac{2\pi}{a}\left(-\frac{\sqrt{3}}{9},\,
\frac{1}{3}\right).
\end{align}
For momenta near the $K$-points, $\vec{k}=\vec{K}(\vec{K}')+\tilde{\vec{k}}$, we find that $f_{\vec{k}}$ is linear in $\tilde{\vec{k}}$,
i.e., $f_{\vec{k}}=\hbar v_F(\lambda \tilde{k}_x-i\tilde{k}_y)$ with valley index $\lambda=\pm$, where $v_F=3ta/2\hbar$ is the Fermi velocity, and in what follows
we omit for the simplicity of notation the tilde over momentum. Thus, we obtain the low-energy Hamiltonian near the $K(K^{\prime})$-point in the
form [\onlinecite{Malcolm_2016}]
\begin{align}\label{Hd-hamiltonian}
&\mathcal{H}_{\lambda}=\hbar v_F(\lambda S_x k_x+S_yk_y)=\hbar v_F\left(\begin{array}{ccc}
0 & \cos\Theta (\lambda k_{x}-ik_y) & 0\\
\cos\Theta (\lambda k_{x}+ik_y) & 0 & \sin\Theta  (\lambda k_{x}-ik_y)\\
0 & \sin\Theta (\lambda k_{x}+ik_y) & 0
\end{array}\right), \nn
& S_x=\left(\begin{array}{ccc}
0 & \cos\Theta & 0\\
\cos\Theta & 0 & \sin\Theta\\
0 & \sin\Theta & 0
\end{array}\right), \quad S_y=\left(\begin{array}{ccc}
0 & -i\cos\Theta & 0\\
i\cos\Theta & 0 & -i\sin\Theta\\
0 & i\sin\Theta & 0
\end{array}\right),
\end{align}
where $\mathbf{S}$ are the spin matrices of the spin 1 representation. The Hamiltonian acts on
three-component wave functions $\Psi^T=(\Psi_{A},\Psi_{C},\Psi_{B})$. The full Hamiltonian, which includes both valleys, is  given by block-diagonal matrix $\text{diag}(H_{+},H_{-})$ and acts on 6-component spinors $(\Psi_{+},\Psi_{-})^{T}$.

It is straightforward to describe the interaction with a magnetic field via the standard Peierls substitution
$\vec{k}\to \vec{k}+\frac{e}{\hbar c}\vec{A}$ in the Hamiltonian. In the following we will use the freedom of the choice of the gauge of vector
potential $\vec{A}$ in order to simplify calculations in particular geometries.

Generally the boundary conditions can be determined from the requirement of the absence of current through the sample boundary. In our case the current in direction $\vec{n}$ for the $K(K')$ valley is given by matrix $ \vec{n}\vec{J}_{\lambda}=\lambda S_x n_x+S_y n_y$. The boundary condition for the wave function of electron states at the boundary is then defined as the requirement that the matrix element of electric current normal to the boundary vanishes, i.e., $\la\Psi_B\right|(\vec{J}_{+}+\vec{J}_{-})\vec{n}\left|\Psi_B\right\rangle=0$. \\

\subsection{Solutions in infinite lattice}
\label{sec:infinite}

For infinite sample we choose the gauge in the form $A = (-By, 0)$, which preserves
translational invariance in the $x$ direction. Then we seek wave functions in the form
$\Psi_\mu = e^{i k_x x}\psi_\mu$.
The Schr\"{o}dinger equation reads
\begin{align}\label{eq:free_system}
\begin{pmatrix}
0 & \cos{\Theta} (\lambda (k_x l - y/l) - l \partial_y) & 0 \\
\cos\Theta(\lambda (k_x l - y/l) + l \partial_y) & 0 & \sin\Theta(\lambda (k_x l - y/l) - l \partial_y) \\
0 & \sin\Theta(\lambda (k_x l - y/l) + l\partial_y) & 0
\end{pmatrix}
\begin{pmatrix}
\psi_A \\
\psi_C \\
\psi_B
\end{pmatrix}
= \frac{\widetilde{\epsilon}}{\sqrt{2}}
\begin{pmatrix}
\psi_A \\
\psi_C \\
\psi_B
\end{pmatrix},
\end{align}
where $\widetilde{\epsilon} = \frac{2\epsilon}{\epsilon_0}$, $ l = \sqrt{\hbar c/|eB|}$ is the magnetic length, and
$\epsilon_0 = \sqrt{2\hbar v_F^2 |eB|/c}$ is the Landau energy scale. It is convenient to introduce new variable $\xi = k_x l - y/l$.
The first and third lines of the system define $\psi_A$ and $\psi_B$ in terms of $\psi_C$ in the case $\widetilde{\epsilon}\neq 0$
\begin{align}
\psi_{A} = \sqrt{2}\cos\Theta \frac{\lambda\xi + \partial_\xi}{\widetilde{\epsilon}} \psi_{C},\quad
\psi_{B} = \sqrt{2}\sin\Theta \frac{\lambda\xi - \partial_\xi}{\widetilde{\epsilon}} \psi_{C}.
\end{align}
Substituting them into the second line of system \eqref{eq:free_system}, we find the following second-order equation for $\psi_{C}$:
\begin{align}\label{eq:eq_on_psi_C}
(\partial^2_\xi - \xi^2)\psi_C + \left(\lambda \cos 2\Theta + \frac{\widetilde{\epsilon}^2}{2}\right)\psi_C = 0.
\end{align}
The general solution of this equation can be written in terms of the parabolic cylinder functions $U$ and $V$ [\onlinecite{Abramowitz}]
\begin{align}\label{eq:zigzag_general}
\psi_C(y) = C_1U\left(-\frac{\widetilde{\epsilon}^2}{4} - \frac{\lambda \cos 2\Theta}{2},\sqrt{2}\xi\right) + C_2V\left(-\frac{\widetilde{\epsilon}^2}{4} - \frac{\lambda \cos 2\Theta}{2},\sqrt{2}\xi\right),
\end{align}
where $C_1$ and $C_2$ are arbitrary constants.
This solution is finite at large distances $|\xi|\to\infty$ only when $C_2 = 0$ and the index of $U$ function is $-\frac{\widetilde{\epsilon}^2}{4} - \frac{\lambda \cos 2\Theta}{2} = -n - 1/2$, where $n=0,\,1,\,\dots$ are nonnegative integers. From this condition we obtain the energy spectrum
\begin{align}\label{eq:free_spectrum}
\epsilon_{n}(\Theta) = \pm \epsilon_0\sqrt{n + 1/2(1 - \lambda \cos 2 \Theta)},
\end{align}
which coincides with the results obtained in Refs.[\onlinecite{Bercioux,Raoux}] .
One should note that for $\Theta\neq 0,\frac{\pi}{4}$ the spectrum is different in the $K$ and $K'$ valleys. The corresponding wave
functions can be written in terms of Hermite polynomials,
\begin{align}
	\Psi_{+}=Ne^{ik_x x-\xi^2/2}
	\begin{pmatrix}
	\frac{2^{1/2}n\cos\Theta}{\sqrt{n + \sin^2\Theta}}H_{n - 1}(\xi)\\
	H_{n}(\xi)\\
	\frac{2^{-1/2}\sin\Theta}{\sqrt{n + \sin^2\Theta}}H_{n + 1}(\xi)
	\end{pmatrix},\quad
	\Psi_{-}=Ne^{ik_x x-\xi^2/2}
	\begin{pmatrix}
	-\frac{2^{-1/2}\cos\Theta}{\sqrt{n + \cos^2\Theta}}H_{n + 1}(\xi)\\
	H_{n}(\xi)\\
	-\frac{2^{1/2}n\sin\Theta}{\sqrt{n + \cos^2\Theta}}H_{n - 1}(\xi)
	\end{pmatrix},
\end{align}
where $N = \frac{1}{\sqrt{2\pi l}}\frac{1}{\sqrt{2^{n + 1}n!\sqrt{\pi}}}$ is the normalization constant and $H_{-1}(\xi)=0$ by definition.

The solutions above were derived for nonzero energy $\widetilde{\epsilon}\neq 0$. Therefore, we should analyze the case
$\widetilde{\epsilon}=0$ separately. For the electron states of zero energy, system \eqref{eq:free_system} reduces to
\begin{align}\label{eq:zero_band_free}
\cos\Theta(\lambda\xi + \partial_\xi)\psi_C = 0,\quad \cos\Theta(\lambda\xi - \partial_\xi)\psi_A + \sin\Theta(\lambda\xi+\partial_\xi)\psi_B = 0,\quad \sin\Theta(\lambda\xi - \partial_\xi)\psi_C = 0.
\end{align}
The first and third equations immediately give $\psi_{C}=0$, and we are left with one equation for two functions $\psi_A$ and $\psi_B$. 
This reflects the fact that the Hamiltonian remains singular (the determinant of the free Hamiltonian (\ref{TB-Hamiltonian}) is 
zero) even in the presence of magnetic field [\onlinecite{Yuce}]. As was noted in Ref.[\onlinecite{Bercioux}], there are two possible types of solutions: the 
conventional zero Landau level and a solution specific to the $\alpha-\mathcal{T}_3$ model. The first type is characterized by single 
nonvanishing component
\begin{align}
	\Psi_{+}=N_0\left(0,\,0,\, e^{ik_x x}e^{-\xi^2/2}\right)^{T},\quad	\Psi_{-}=N_0\left(e^{ik_x x}e^{-\xi^2/2},\,0,\,0\right)^{T},
\end{align}
with $N_0=\frac{1}{\sqrt{2\pi l\sqrt{\pi}}}$, while the second one has two nonvanishing components
\begin{align}
\Psi_{+} =
N e^{ik_x x -\xi^2/2}
\begin{pmatrix}
-\frac{2(n + 1)\sin\Theta}{\sqrt{n + \sin^2\Theta}} H_{n - 1}(\xi) \\
0\\
\frac{\cos\Theta}{\sqrt{n + \sin^2\Theta}} H_{n + 1}(\xi)
\end{pmatrix} \quad
\Psi_{-} =
N e^{ik_x x -\xi^2/2}
\begin{pmatrix}
\frac{\sin\Theta}{\sqrt{n + \cos^2\Theta}} H_{n + 1}(\xi) \\
0\\
-\frac{2(n + 1)\cos\Theta}{\sqrt{n + \cos^2\Theta}} H_{n - 1}(\xi)
\end{pmatrix},
\label{general-solution-model}
\end{align}
where $n = 1, 2,...$ and $N = \frac{1}{\sqrt{2\pi l\sqrt{\pi}}\sqrt{2^{n + 1} (n-1)!(n +1)}}$. Note that the conventional graphene-like Landau level formally coincides with topological solution
(\ref{general-solution-model}) with $n=-1$ and $H_{-2}(\xi)=0$.

\section{Zigzag termination}
\label{sec:zigzag}

\begin{figure}
	\includegraphics[scale=0.5]{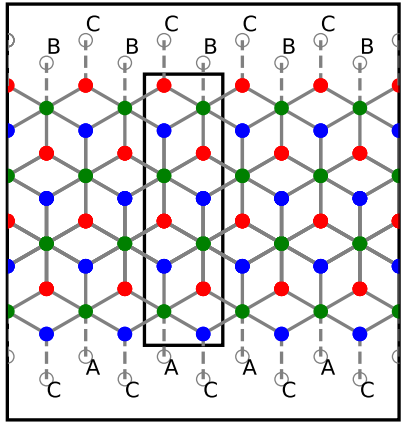}
	\caption{An example of terminated $\alpha-{\cal T}_3$ lattice with zigzag boundary is shown. The black rectangle shows cell for which the tight-binding calculations were performed.}
	\label{fig2}
\end{figure}
In the case of zigzag termination shown in Fig.\ref{fig2}, the translational invariance is preserved only along the $x$ direction. Thus, we can use the general solution \eqref{eq:zigzag_general} with appropriate boundary conditions to find the spectrum for different types of terminations. Since zigzag conditions do not mix different valleys, we can perform calculations for one valley.

The requirement of vanishing of the normal current at the boundary (in the $\vec{n}=(0,\pm 1)$ directions) implies the following
equation for boundary wave functions:
\begin{align}
\la\Psi_B\right|\vec{J}{\vec{n}}\left|\Psi_B\right\rangle =i \psi_C^*(\psi_A \cos\Theta - \sin\Theta \psi_B) +i \psi_C(\sin\Theta\psi_B^* - \cos\Theta\psi_A^*) = 0.
\end{align}
The two main possible solutions are $\psi_C\bigg|_{B} = 0$  and $(\psi_A \cos\Theta - \psi_B \sin\Theta)\bigg|_{B} = 0$. As was found in
[\onlinecite{Oriekhov}], the first condition corresponds to the C, AC or BC types of termination at low energies, while the last one corresponds
to the AB termination. Using relations between the A, B and C components, the termination of the AB type gives
$\psi_C' + \xi \cos2\Theta\psi_C = 0$. In the same way for the termination of the C type we obtain
$\sin\Theta\psi_A+\cos\Theta\psi_B=0$.

Clearly, there are 3 main combinations of terminations: C-C, C-AB, and BA-AB. Below we analyze all three types and determine the main features of the corresponding spectra.

\subsection{The \textbf{C}-\textbf{C} boundary conditions.}

\begin{figure}
\centering
\includegraphics[scale=0.65]{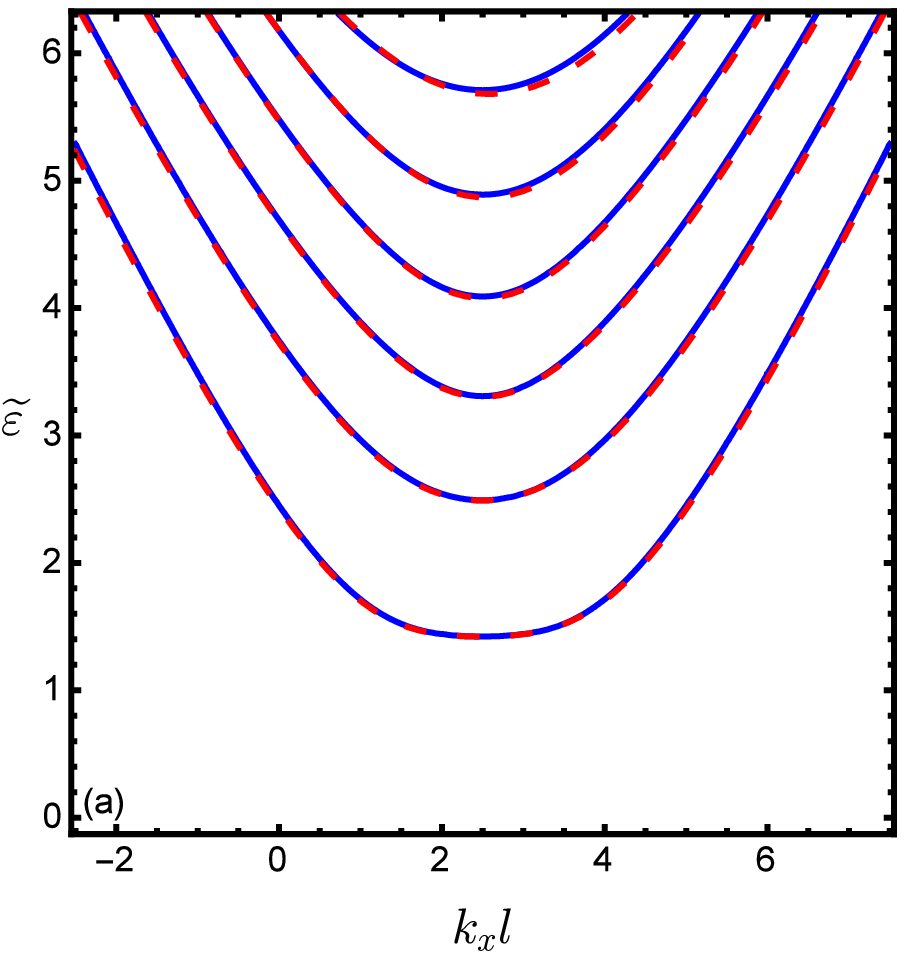}\quad\quad
\includegraphics[scale=0.65]{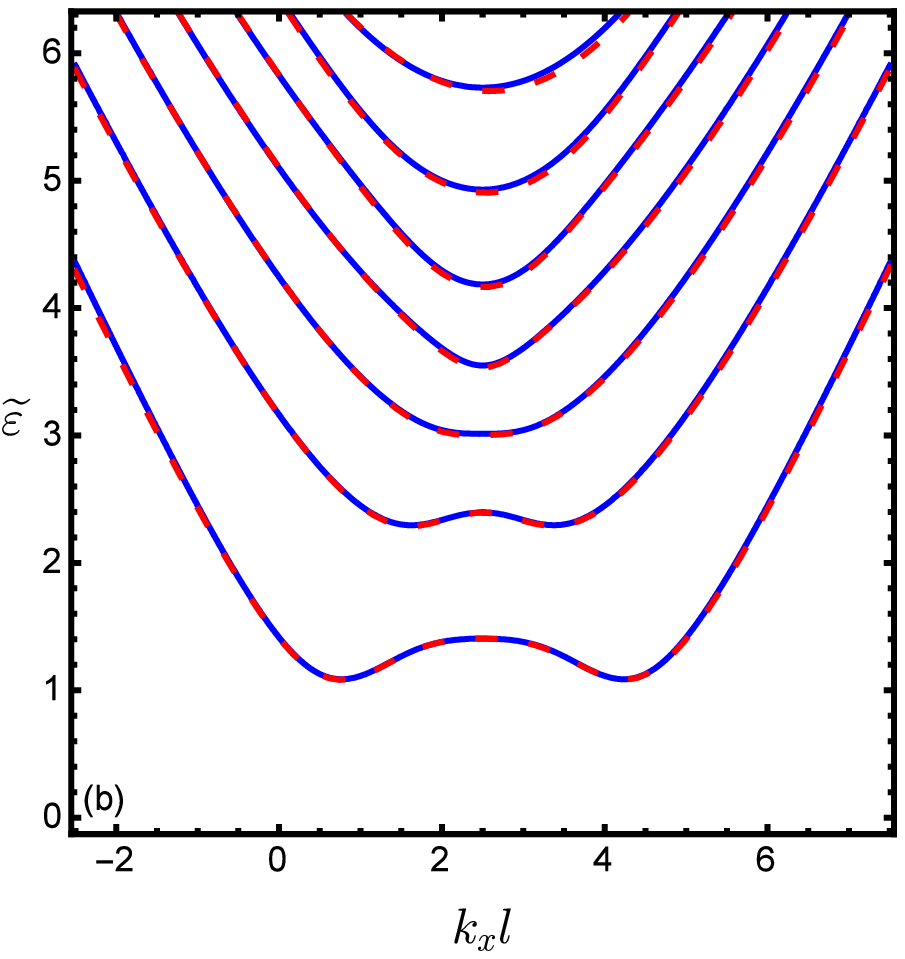}
\includegraphics[scale=0.65]{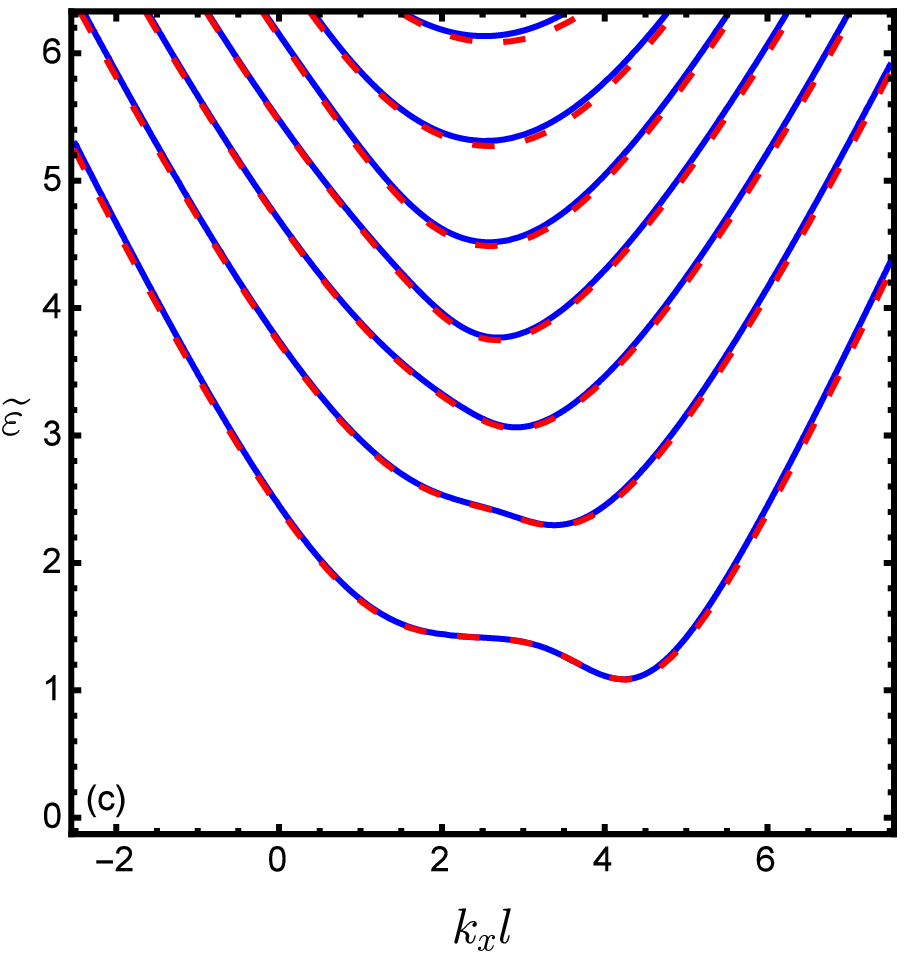}\quad\quad
\includegraphics[scale=0.65]{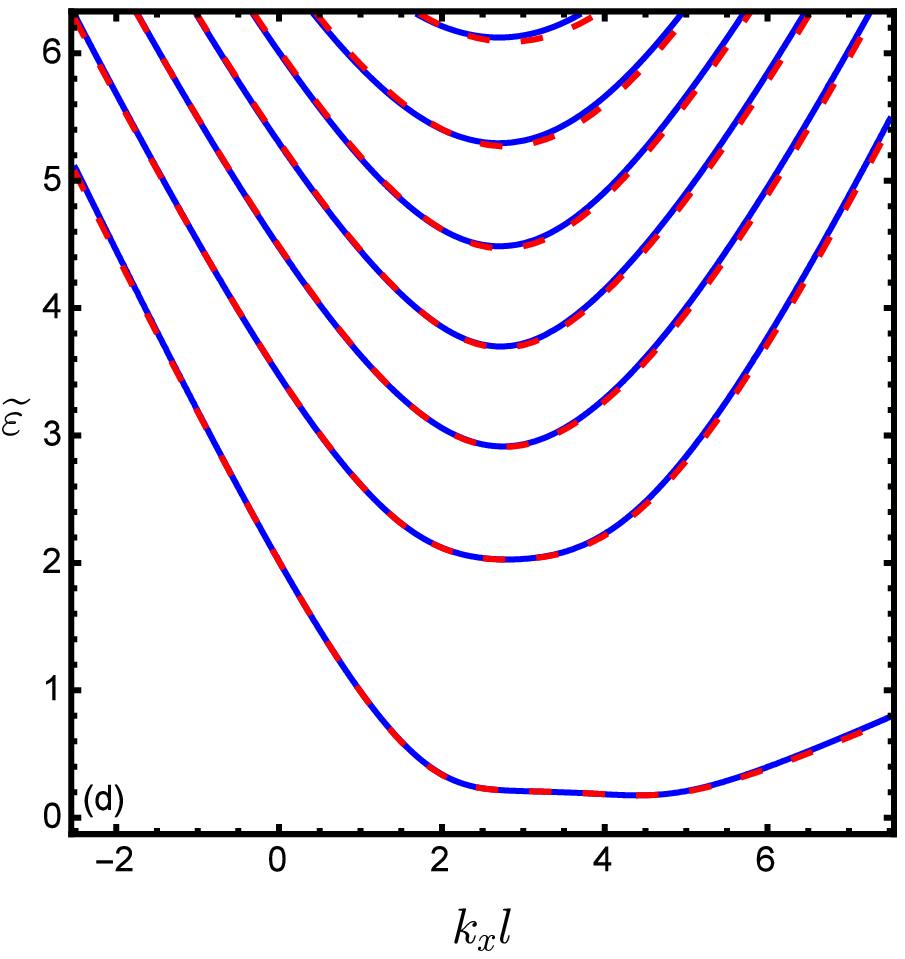}
\caption{The energy spectrum of analytic solutions (blue solid lines) for zigzag terminations of (panel a) C-C type, (panel b)
	BA-AB type, (panel c) C-AB type for the strip of the width $L/l=5$ and parametric angle $\Theta=\frac{\pi}{4}$. We also plot solutions for
	the small angle value $\Theta=\frac{\pi}{30}$ (panel d) which demonstrate the transformation of the lowest dispersive level into 
	the localized edge state. The results of numerical tight-binding calculations
	(red dashed lines) are in good agreement with analytical curves. The number of atoms in calculation cell was taken approximately 700. The 
	chosen angle $\Theta=\frac{\pi}{4}$
	corresponds to dice model.}
	\label{fig:zigzag_solution}
\end{figure}

Let us, for definiteness, consider the C-C case.
Substituting the general solution \eqref{eq:zigzag_general} into conditions $\psi_C(y=0)=\psi_C(y=L)=0$, we find the following
system of equations for constants $C_1$ and $C_2$:
\begin{align}
	C_1U\left(-\frac{\widetilde{\epsilon}^2}{4} - \frac{\lambda \cos 2\Theta}{2},\sqrt{2}\xi_{j}\right) + C_2V\left(-\frac{\widetilde{\epsilon}^2}{4} - \frac{\lambda \cos 2\Theta}{2},\sqrt{2}\xi_{j}\right)=0,\quad j=1,\,2,
\end{align}
where $\xi_{1}=k_x l$ and $\xi_{2}=k_x l - L / l$.
The corresponding characteristic equation in K valley reads
\begin{align}\label{eq:CC-characteristic}
&U\left(-\frac{\widetilde{\epsilon}^2}{4} - \frac{\cos 2\Theta}{2},\sqrt{2}k_x l\right)V\left(-\frac{\widetilde{\epsilon}^2}{4} - \frac{\cos 2\Theta}{2},\sqrt{2}(k_x-k_0)l\right) - \nn
&U\left(-\frac{\widetilde{\epsilon}^2}{4} - \frac{\cos 2\Theta}{2},\sqrt{2}(k_x-k_0)l\right)V\left(-\frac{\widetilde{\epsilon}^2}{4} - \frac{\cos 2\Theta}{2},\sqrt{2}k_x l\right) = 0,
\end{align}
where $k_0 = L/l^2$ is determined by the width of ribbon. Using formulas 19.4.2 and 19.4.3 in [\onlinecite{Abramowitz}], we find that this equation is symmetric with respect to $k_0/2$, i.e., $\widetilde{\epsilon}(k_x)=\widetilde{\epsilon}(k_0-k_x)$. Solving equations numerically we determine the dependence of dimensionless energy
parameter $\widetilde{\epsilon}$ on the wave vector $k$. To compare analytical results with numerical tight-binding calculations, we
use the Harper equation approach (for more details, see Refs.[\onlinecite{Chao-Zhang,Chen_nanoribbons}]). The results of two methods
are shown for several lowest levels with nonzero energy $\widetilde{\epsilon}\neq 0$ in Fig.\ref{fig:zigzag_solution}.

Similar to Ref.[\onlinecite{Heuser}], we find analytical expressions for bulk and edge electron states in the limit of wide strip $L/l\to\infty$ (strong magnetic field). Using the asymptotic expansions \eqref{eq:asympxU} and \eqref{eq:asympxV}
	we find the following $k_0\to+\infty$ limit of the characteristic equation:
	\begin{align}\label{eq:char_approx}
	\frac{U(a,\sqrt{2}k_x l)}{V(a,\sqrt{2}k_x l)}=\frac{\pi}{\Gamma(a+1/2)\sin(\pi a)}.
	\end{align}
Applying Eq.\eqref{eq:U_minus_x}, we can rewrite this equation in the "half-plane" form $U(a,-\sqrt{2}k_x l)=0$.
Expanding this equation to linear order in $\sqrt{2}k_x l$, we find the edge electron spectrum. Substituting $a=-2n-\frac{3}{2}+\delta_n$ with small $\delta_n$ and integer $n$, we obtain in the leading order
\begin{align}
	\delta_n =\frac{4\Gamma(n+3/2)}{\pi n!}k_x l.
\end{align}

The corresponding energy levels are given by
\begin{align}\label{eq:c_edge}
	\epsilon_{n}(\Theta)=\pm \epsilon_0\sqrt{2n+\frac{3-\lambda\cos(2\Theta)}{2}}\left(1-\frac{4\Gamma(n+3/2)}{\pi n! (4n+3-\lambda\cos(2\Theta))}k_x l\right).
\end{align}
The second term in brackets represents small correction from $k_x$. One should note that the main term scales as $\sqrt{2 n}$, comparing to the infinite system spectrum \eqref{eq:free_spectrum}, which scales as $\sqrt{n}$. By using Eq.\eqref{eq:char_approx}, one can also find the approximate spectrum of bulk electron states, which corresponds to the limit $k_x l\gg 1$ and $k_x\ll k_0$. Using asymptotic expansions for positive arguments, we obtain
\begin{align}
	\sin(\pi a)\Gamma(a+1/2)e^{-k_x^2 l^2}(\sqrt{2}k_x l)^{-2a}=\sqrt{2\pi}.
\end{align}
We seek solution for the $a=-n-\frac{1}{2}+\delta_n$, with integer $n\geq 0$ and small $\delta_n$. Expanding this equation to leading order in $\delta$, we find $\delta_n=-\frac{e^{-k_x^2 l^2} (\sqrt{2}k_x l)^{2 n+1}}{\sqrt{2 \pi } n!}$. This results in the following correction to spectrum:
\begin{align}\label{eq:c_bulk}
    \epsilon_{n}(\Theta)=\pm \epsilon_0\sqrt{n+\frac{1-\lambda\cos(2\Theta)}{2}}\left(1+\frac{e^{-k_x^2 l^2} (\sqrt{2}k_x l)^{2 n+1}}{\sqrt{2 \pi } n!(2n+1-\lambda\cos(2\Theta))}\right).
\end{align}
In the limit $k_x l\to \infty$ we recover the infinite system spectrum \eqref{eq:free_spectrum}. The positive sign of the second term in brackets indicates that the global minima of spectra is situated deep in the bulk. This agrees with panel a of the Fig.\ref{fig:zigzag_solution}.

When the width of ribbon is 3 or 4 times smaller than the magnetic length, we find that the spectra have little overlap with the free
bulk spectra \eqref{eq:free_spectrum}. Additionally, the separation between the nearest levels quickly diminish with the decrease
of $L$. When the ribbon's width is 6 or 7 times larger than the magnetic length, nearly flat plateaus are formed on the lowest
levels around the central wave number $k = k_0/2$. We also find that for lower levels the width of plateaus in the $k$ space is larger. These
features are similar to those obtained in graphene [\onlinecite{Gusynin}].

The ribbon with C-C boundary conditions has insulating type of spectrum even without applied magnetic field 
[\onlinecite{Chen_nanoribbons,Oriekhov}]. In order to analyze the spectral gap behavior as a function of weak magnetic field $l\gg L$, we need to 
consider the asymptotics of parabolic cylinder functions $U(a,x), V(a,x)$ at $|a| \gg x^2$, because in our case
$|a| \sim B^{-1}$ (and $a<0$) and $x \sim B^{1/2}$. With the help of Eqs.\eqref{eq:largea1}-\eqref{eq:largea4}, the equation \eqref{eq:CC-characteristic} reduces to
\begin{align}
\tilde{A}\sin\left[\frac{k_0 l(k_0^2 l^2 - 12(\widetilde{\epsilon}^2+2\cos 2\Theta))}{12\sqrt{2}\sqrt{\widetilde{\epsilon}^2+2\cos 2\Theta}}\right] = 0,
\end{align}
where $\tilde{A}\neq 0$ is a coefficient of expansion.
For the lowest Landau level in the upper band the argument of the sine equals $-\pi$. Thus, introducing the spectral gap $\Delta$ as 
the energy distance between the lowest level of the positive energy band and the zero energy band, we find 
its following approximate expression in the weak magnetic field limit
\begin{align}
\Delta=\frac{\hbar  v_F}{L}\left(\pi-\frac{\cos (2 \Theta )}{2 \pi  }\left(\frac{L}{l}\right)^2+\frac{\pi ^2-3 \cos^2 (2\Theta)}{24 \pi^3 }\left(\frac{L}{l}\right)^4\right).
\end{align}
One should note that the linear in $B$ correction is negative, and it vanishes in the dice model $\Theta=\frac{\pi}{4}$. 

The results for other types of terminations are discussed in Appendix \ref{appendix:add_zigzag}. Note that magnetic field opens a gap for all types of termination. However, the shapes of the spectrum are different in each case (see Fig.\ref{fig:zigzag_solution}).
\subsection{Zero energy band}

\begin{figure}
	\includegraphics[width=0.32\textwidth]{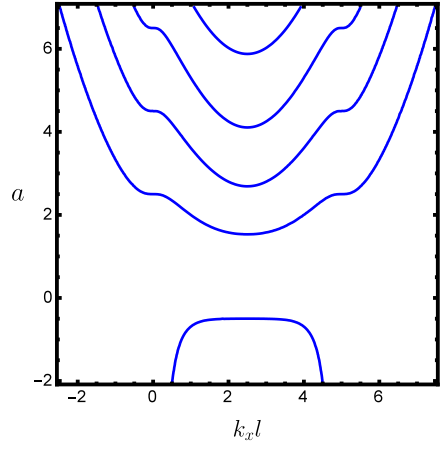}
	\caption{The analytic solutions of Eq.\eqref{eq:zigzag_zeroband} for the zero energy band and the BA-AB zigzag boundary 
	conditions at $L/l = 5$ and $\Theta=\pi/4$. }
	\label{fig:AB_flat_band}
\end{figure}

In the case of zero energy, there is only one middle equation of system \eqref{eq:zero_band_free} for two components $A$ 
and $B$. The third component $\psi_{C}\equiv0$ everywhere. The corresponding wave
functions can be searched in the following general form:
\begin{align}\label{eq:AB_ansatz}
&\psi_A = (\xi + \partial_\xi)(C_1 U(-a, \sqrt{2}\xi) + C_2V(-a,\sqrt{2}\xi)),\quad
\psi_B = (\xi - \partial_\xi)(C_3 U(-a, \sqrt{2}\xi) + C_4V(-a,\sqrt{2}\xi)),
\end{align}
with arbitrary real parameter $a$ and constants $C_1,\,\dots C_{4}$.
Inserting them into second equation of system \eqref{eq:zero_band_free}
\begin{align}
U(-a, \sqrt{2}\xi)(\cos\Theta(1/2-a )C_1 - \sin\Theta(1/2+a)C_3) + V(-a, \sqrt{2}\xi)(\cos\Theta(1/2-a)C_2 - \sin\Theta(1/2+a)C_3) = 0,
\end{align}
we obtain linear relations between coefficients
$C_1 = C_3\tan\Theta \frac{1/2+a}{1/2-a},\,
C_2 = C_4\tan\Theta \frac{1/2+a}{1/2-a}$. Substituting them into the expressions \eqref{eq:AB_ansatz} for $\psi_A$ and $\psi_B$,
we find
\begin{align}
&\psi_A = \sqrt{2}\sin\Theta\bigg[-A(a + 1/2)U(1-a, \sqrt{2}\xi) + B V(1-a, \sqrt{2}\xi)\bigg], \\
&\psi_B = \sqrt{2}\cos\Theta\bigg[A U(-a - 1, \sqrt{2}\xi) + B(1/2 - a)V(-a - 1, \sqrt{2}\xi)\bigg].
\end{align}

For example, in the case of the BA-AB termination, we have the characteristic equation
\begin{align}\label{eq:zigzag_zeroband}
\text{det}\begin{vmatrix}
-(a + 1/2)U(-a + 1, \sqrt{2}\xi_1) - U(-a - 1, \sqrt{2}\xi_1) &  V(-a + 1, \sqrt{2}\xi_1) - (1/2 - a)V(-a - 1, \sqrt{2}\xi_1)\\
-(a + 1/2)U(-a + 1, \sqrt{2}\xi_2) - U(-a - 1, \sqrt{2}\xi_2) &  V(-a + 1, \sqrt{2}\xi_2) - (1/2 - a)V(-a - 1, \sqrt{2}\xi_2)\\
\end{vmatrix} = 0,
\end{align}
where $\xi_1 = k_xl$ and $\xi_2 = k_xl - L/l$. The numerical solutions for parameter $a$ as a function of $kl$ are given in
Fig.\ref{fig:AB_flat_band}. For other types of boundary conditions the calculations are similar, and we found that there are
solutions in all cases for every $kl$. This finding agrees with the results of Ref.[\onlinecite{Chen_nanoribbons}] that the
flat band does survive even in the presence of a magnetic field and boundaries for $\Theta\neq 0$.

Notably,
	there is a solution with negative values of $a$, which corresponds to graphene-like zero Landau level, for which one nonzero component is
exponentially suppressed compared to the second component. The main difference from graphene is that such a solution is
dispersionless.

\section{Armchair termination}
\label{sec:armchair}

\begin{figure}
	\includegraphics[scale=0.4]{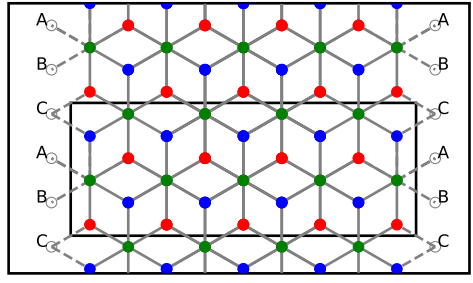}
	\caption{An example of terminated $\alpha-{\cal T}_3$ lattice with armchair boundary is shown. The black rectangle shows cell for which the 
     tight-binding calculations were performed.}
	\label{fig3}
\end{figure}

\begin{figure}
	\centering
	\includegraphics[width=0.3\textwidth]{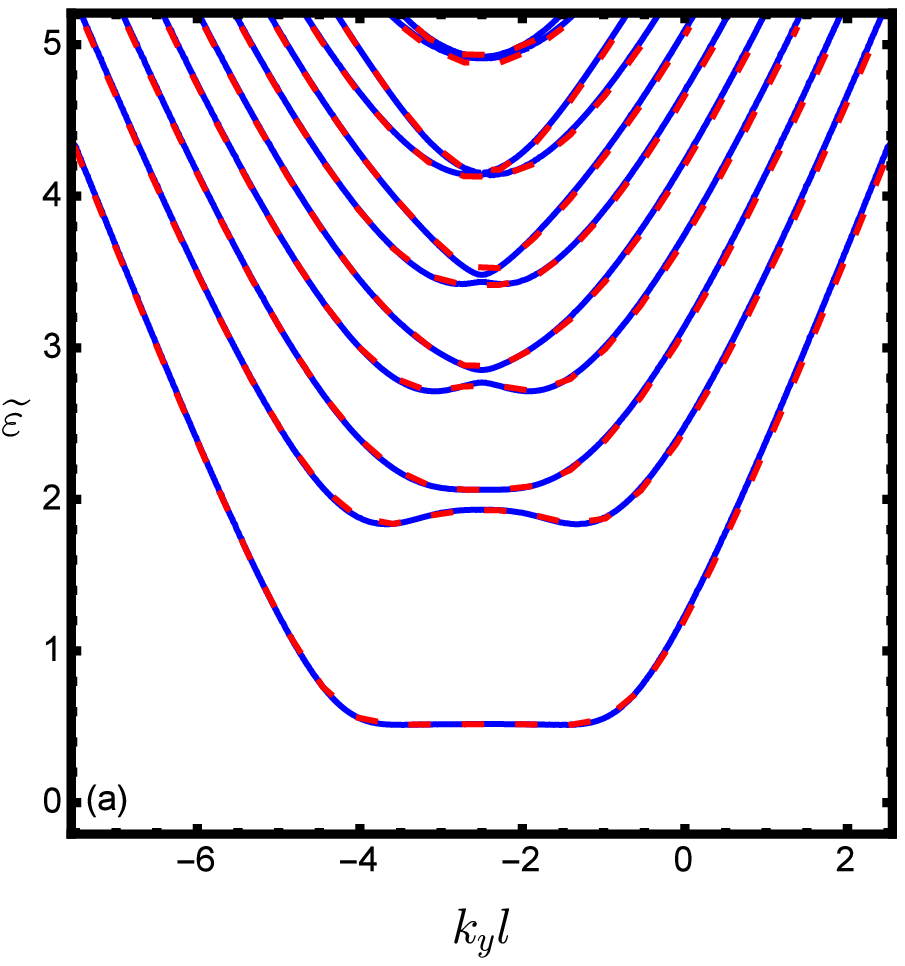}\quad
	\includegraphics[width=0.3\textwidth]{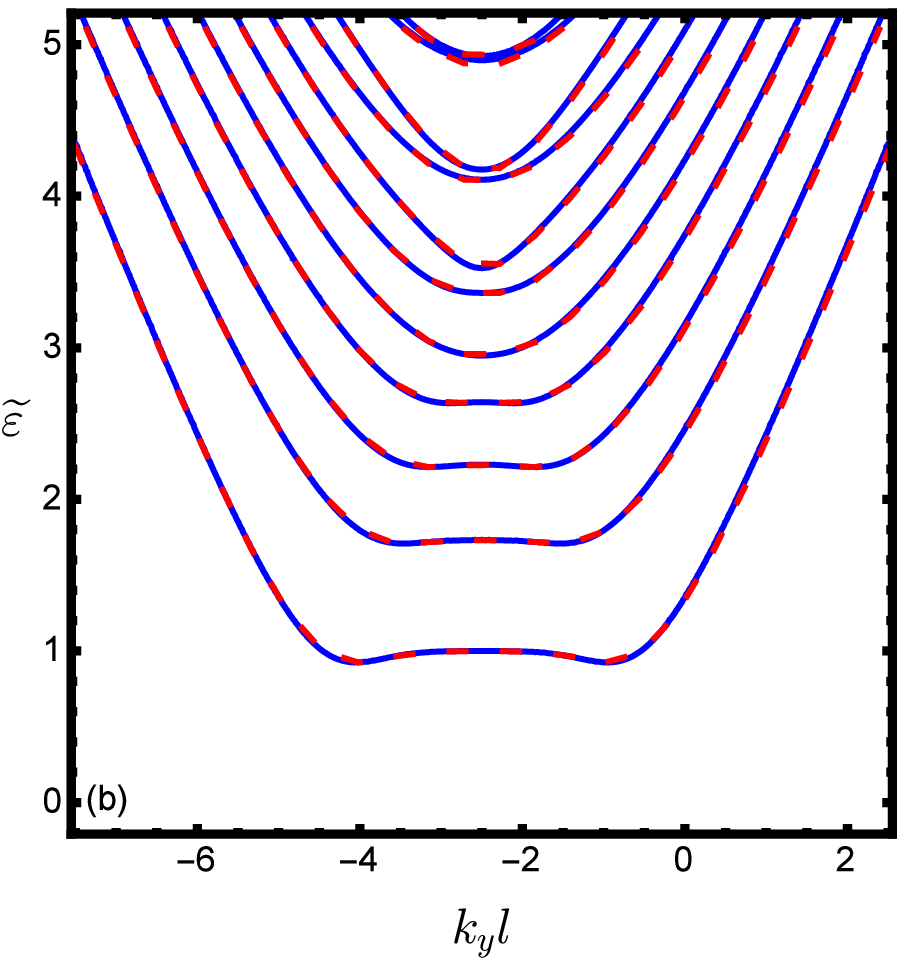}\quad
	\includegraphics[width=0.3\textwidth]{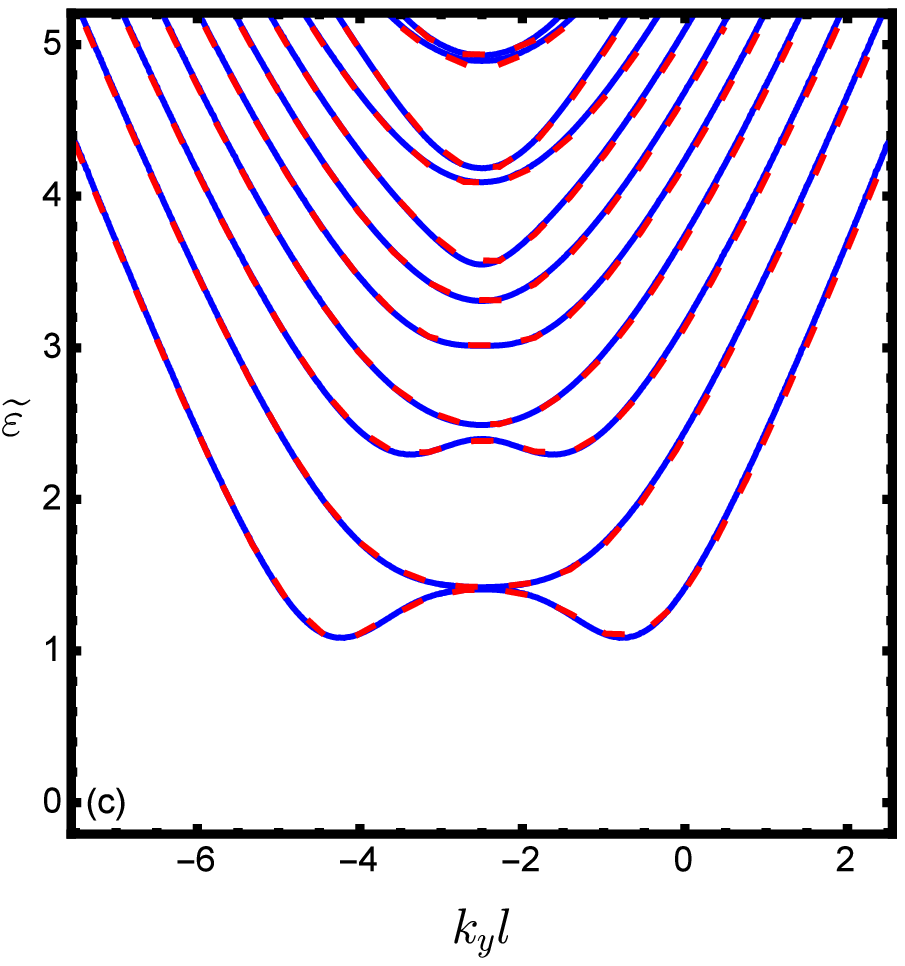}
	\caption{The energy spectrum of analytic solutions (blue solid lines) for armchair terminations with parametric angle $\Theta=\frac{\pi}{12}
		$ (panel a), $\Theta=\frac{\pi}{6}$ (panel b) and $\Theta=\frac{\pi}{4}$ (panel c). The width of the ribbon is $L/l = 5$. Red dashed lines 
		represent the results of tight-binding calculation with 251 atomic rows. }
	\label{fig:armchair_compare}
\end{figure}

In order to consider the armchair boundary condition along the $y$ direction (see Fig.\ref{fig3}), it is convenient to
use the vector potential in the form $A = (0, Bx)$. This choice preserves translational invariance in the $y$ direction, therefore,
we can seek wave functions in the form $\Psi=e^{ik_y y}\psi(x)$. The eigenstate equation in the K(K') valley has the form
\begin{align}
\hbar v_F
\begin{pmatrix}
0 & \cos\Theta(-i\lambda\partial_x - ik_y - \frac{ieBx}{\hbar c}) & 0\\
\cos\Theta(-i\lambda \partial_x + ik_y + \frac{ieBx}{\hbar c}) & 0 & \sin\Theta(-i\lambda\partial_x - ik_y - \frac{ieBx}{\hbar c})\\
0& \sin\Theta(-i\lambda\partial_x + i k_y + \frac{ieBx}{\hbar c}) & 0
\end{pmatrix}
\begin{pmatrix}
\psi_A \\ \psi_C \\ \psi_B
\end{pmatrix}
=
\epsilon
\begin{pmatrix}
\psi_A \\ \psi_C \\ \psi_B
\end{pmatrix}.
\end{align}
Introducing new variable $\xi = k_y l + x/l$ one can rewrite equations as follows:
\begin{align}\label{eq:hamiltonian_armchair}
i
\begin{pmatrix}
0 & \cos\Theta(-\lambda\partial_\xi - \xi) & 0\\
\cos\Theta(-\lambda\partial_\xi + \xi) & 0 & \sin\Theta(-\lambda\partial_\xi - \xi)\\
0& \sin\Theta(-\lambda\partial_\xi + \xi) & 0
\end{pmatrix}
\begin{pmatrix}
\psi_A \\ \psi_C \\ \psi_B
\end{pmatrix}
=
\frac{\widetilde{\epsilon}}{\sqrt{2}}
\begin{pmatrix}
\psi_A \\ \psi_C \\ \psi_B
\end{pmatrix}.
\end{align}
In each valley this system reduces to the following second-order equation for the $\phi_C$ component:
\begin{align}
(\p_{\xi}^{2}-\xi^2) \psi_C + \left(\lambda\cos{2\Theta} + \frac{\widetilde{\epsilon}^2}{2}\right)\psi_C = 0,
\end{align}
which coincides with Eq.\eqref{eq:eq_on_psi_C}.
Since the normal electric current at boundaries $x=0$ and $x=L$ should vanish,
\begin{align}
&\la\Psi_B\right|\vec{J}\vec{n}\left|\Psi_B\right\rangle = \psi_C(\psi_B^*\sin\Theta + \psi_A^* \cos\Theta) + \psi_C^*(\psi_A \cos\Theta
+ \psi_B \sin\Theta) -\nn
&\psi_{C'}(\psi_{A'}^* \cos\Theta + \psi_{B'}^* \sin\Theta) - \psi_{C'}^*(\psi_{B'} \sin\Theta + \psi_{A'} \cos\Theta),
\end{align}
we find the following armchair boundary conditions [\onlinecite{Akhmerov,Brey,Oriekhov}] ($\mu=A,\,B,\,C$):
\begin{align}\label{eq:armchair_conditions}
\psi_{\mu}(x=0)=\psi_{\mu^{\prime}}(x=0),\quad \psi_{\mu}(x = L) = e^{i \Delta KL}\psi_{\mu^{\prime}}(x = L).
\end{align}
By using system \eqref{eq:hamiltonian_armchair} we rewrite these 6 conditions as 4 nontrivial conditions for the $\psi_C$ components
\begin{align}
\psi_C' = -\psi_{C'}'\bigg|_{x=0}\bigg.,\quad
\psi_C = \psi_{C'}\bigg|_{x=0}\bigg. ,\quad \psi_{C'}' = -e^{i \Delta KL}\psi_{C'}'\bigg|_{x=L}\bigg.,\quad \psi_C= e^{i \Delta KL}\psi_{C'} 
\bigg|_{x=L}\bigg.
\end{align}
Since the spectrum practically does not change with the change of $\Delta KL$, only the case of $\cos\Delta KL = 1$ is examined.
The results of analytical solutions (see Appendix \ref{appendix:armchair_add} for detailed derivations) are presented in Fig.\ref{fig:armchair_compare} and compared with those of numerical
tight-binding calculations. One can observe that the magnetic field opens a gap in the spectrum even for the metallic armchair termination. This 
finding agrees with that obtained in Ref.[\onlinecite{Chen_nanoribbons}]. 

\begin{figure}
\includegraphics[width=0.32\textwidth]{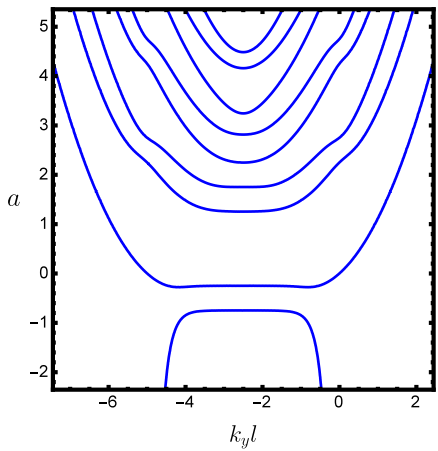}
	\caption{The analytic solutions for the zero energy band in the case of armchair termination for $L/l = 5$. The parameter angle equals
	$\Theta=\frac{\pi}{6}$.}
	\label{fig:armchair_zeroband}
\end{figure}

Using similar method as in zigzag termination case we find a set of zero-energy solutions (see Appendix \ref{appendix:armchair_add}). The corresponding results are shown on Fig.\ref{fig:armchair_zeroband}. Notably,
	there is a solution with negative values of parameter $a$, which corresponds to graphene-like zero Landau level. However, in case of the
	$\alpha-\mathcal{T}_3$ model, such a solution is dispersionless.
 
%Notably,there is a solution with negative values of $a$, which corresponds to graphene-like zero Landau level. However, in case of the $\alpha-\mathcal{T}_3$ model, such a solution is dispersionless.

\section{Summary}
\label{sec:summary}

We studied the electron states in ribbons of the $\alpha-\mathcal{T}_{3}$ lattice with different types of zigzag and armchair 
terminations in a magnetic field. By making use of the low-energy effective theory of pseudospin-1 fermions, we found analytical 
expressions for wave functions of the lowest energy levels and determined the corresponding energy spectrum. Our results agree with 
the results of numerical tight-binding calculations as well as the recent conclusions in [\onlinecite{Chen_nanoribbons}] that the magnetic field 
opens a gap in the spectrum for all types of terminations and that the zero energy flat band survives.  The 
underlying reason of the stability of the flat band is a singular nature of the Hamiltonian of pseudospin-1 fermions. The 
presence of boundaries and magnetic field do not remove this singularity, therefore, the flat band survives.
This is in contrast to the cases where an external potential makes Hamiltonian non-singular and completely destroys the flat band 
(for example, see discussion in our recent paper [\onlinecite{coulomb_alphaT3}] about the influence of the Coulomb potential).

For zigzag-type terminations we found that for symmetric boundary conditions (C-C and BA-AB) the low-energy spectrum near one valley is 
symmetric with respect to the wave vector $L/2l^2$. For asymmetric C-AB boundary conditions the corresponding spectrum is asymmetric, and at low 
values of parametric angle $\Theta$ the sets of localized edge modes are formed. The last finding agrees with the fact that for $\Theta\to0$ (as well as for $\Theta\to\frac{\pi}{2}$) the $\alpha-\mathcal{T}_3$ model describes a graphene-like system. We found simple analytical expressions for the magnitude of the spectral gap at a weak magnetic field, which scales linearly with $B$ and reaches its maximum for the dice model $\Theta=\frac{\pi}{4}$. We also find analytical expressions for spectrum of edge and bulk electron states in the wide strip limit. The energy of corresponding Landau levels is determined by nearest termination type, and scales differently on boundary and deep in the bulk.

For armchair-type ribbons in a magnetic field the gap is present for both metallic and insulating ribbon widths. Thus, the actual number of atomic rows across ribbon can only change the magnitude of gap.

We find that the flat band solutions can be divided into two distinct types. These types
refer to the conventional graphene-like zero Landau level state and a solution specific to the $\alpha-\mathcal{T}_{3}$ model. The
main difference compared to graphene is that both types remain dispersionless even in the presence of boundaries.

\begin{acknowledgements}
	We are grateful to E.V. Gorbar and V.P. Gusynin for fruitful discussions and critical reading of the manuscript.
\end{acknowledgements}

\appendix

\section{Zigzag type ribbons with BA-AB and C-AB boundary conditions}
\label{appendix:add_zigzag}
In this Appendix we present solutions for zigzag ribbons with two other types of terminations: BA-AB and C-AB. 
\subsection{The \textbf{BA}-\textbf{AB} boundary conditions.}
In case of AB-AB termination we have the following characteristic equation
\begin{align}\label{eq:AB_charact}
&\hspace{-0.5em}\left[\sqrt{2} U^{\prime}\left(a, \sqrt{2} k_{x} l\right)+\cos 2 \Theta k_{x} l U\left(a, \sqrt{2} k_{x} l\right)\right]\left[\sqrt{2} V^{\prime}\left(a,\sqrt{2}\left(k_{x}-k_{0}\right) l\right)+\cos 2 \Theta\left(k_{x}-k_{0}\right) l V\left(a,\sqrt{2}\left(k_{x}-k_{0}\right) l\right)\right]-  \nn
&\hspace{-0.5em}\left[\sqrt{2} U^{\prime}\left(a,\sqrt{2}\left(k_{x}-k_{0}\right) l\right)+\cos 2 \Theta\left(k_{x}-k_{0}\right) l U\left(a,\sqrt{2}\left(k_{x}-k_{0}\right) l\right)\right]\left[\sqrt{2} V^{\prime}\left(a,\sqrt{2} k_{x} l\right)+\cos 2 \Theta  k_{x} l V\left(a,\sqrt{2} k_{x} l\right)\right]=0.
\end{align}
The numerical solution are shown on panel b of Fig.\ref{fig:zigzag_solution}. We find that near dispersionless plateaus are formed on the first levels with
$\widetilde{\epsilon}\neq 0$ for $L\gg l$. The spectrum is found to be symmetric with respect to $k_0/2$ wave number.

For $L \gg l$, this equation is simplified to
\begin{align}
\frac{\sqrt{2} U^{\prime}\left(a, \sqrt{2} k_{x} l\right)+\cos 2 \Theta k_{x} l U\left(a, \sqrt{2} k_{x} l\right)}{\sqrt{2} V^{\prime}\left(a,\sqrt{2} k_{x} l\right)+\cos 2 \Theta  k_{x} l V\left(a,\sqrt{2} k_{x} l\right)} = \frac{\pi }{ \Gamma \left(a+\frac{1}{2}\right)\sin \pi  a}
\end{align}
Expanding up to linear order in $\sqrt{2}k_xl$ and substituting $a = -2n - \frac{1}{2} + \delta_n$ (compare with the C-C case where $a=-2n-\frac{3}{2}+\delta_n$), we find the approximate edge electron spectrum
\begin{align}\label{eq:ab_edge}
\epsilon_{n}(\Theta) = \pm \epsilon_0 \sqrt{2 n + \frac{1 - \lambda \cos 2\Theta}{2}} \left(1 - \frac{ \Gamma \left(n+\frac{1}{2}\right)}{\pi  n!}k_x l\right).
\end{align}

For the bulk electron states, i.e., $k_x l \gg 1$, using the asymptotic expansions of parabolic cylinder functions \eqref{eq:asympxU}-\eqref{eq:asymp_derivV}, and searching $a = -n - \frac{1}{2} + \delta_n$, we obtain the following form of the bulk states spectrum
\begin{align}\label{eq:ab_bulk}
\epsilon_{n}(\Theta) = \pm \epsilon_0 \sqrt{n + \frac{1-\lambda  \cos 2 \Theta }{2}} \left(1-\frac {e^{-k_x^2l^2} (\sqrt{2}k_x l)^{2 n+1} \tan ^{2}\Theta }{\sqrt{2 \pi } n! (2 n+1 -\lambda  \cos 2 \Theta )}\right).
\end{align}
While the first term in brackets leads to the same infinite system spectrum \eqref{eq:free_spectrum} as in zigzag C-C case, the second term has opposite sign comparing to that in Eq.\eqref{eq:c_bulk}. This result can be qualitatively understood from the numerical results on panel (b) of Fig.\ref{fig:zigzag_solution}. The spectrum of lowest levels has a form of "mexican hat", with two local minima situated at the edges of near dispersionless plateaus. Thus, the negative correction to free electron spectrum vanishes deep in the bulk, but reaches maximum at some intermediate point.

In the case of BA-AB boundary conditions, which are of metallic type at $B=0$, we find that applied magnetic field opens a gap at all 
values of $\Theta$. In order to analyze the behavior of the spectral gap more quantitatively we consider the limit
$L \rightarrow 0$ or, more precisely, $l \gg L$ that corresponds to a weak applied magnetic field.
We assume that the gap takes minimum value at $k_x = k_0/2$.  Using the Taylor series \eqref{eq:Taylor_expansion1} and \eqref{eq:Taylor_expansion2} for parabolic cylinder functions for small $k_0 l$ and fixed $a$,
we expand Eq.\eqref{eq:AB_charact} at $k_x=k_0/2$ point up to third order in $k_0 l$. The result can be formally written $c_1 k_0l-c_3 (k_0l)^3=0$, where 
$c_1$ and $c_3$ are the Taylor series coefficients. Expanding the corresponding solution $k_0l=\sqrt{c_1/c_3}$ to linear 
order in $\widetilde{\epsilon}$, we find the following approximate value of the energy for the lowest state of the 
upper band: 
\begin{align}
k_0 l = 2\sqrt{-\frac{2 a  +\cos 2\Theta}{4 a \cos 2\Theta + \frac{8a^2 + 1}{3} + \cos^2 2\Theta}} \approx
\frac{\sqrt{6}}{\sin 2\Theta}\widetilde{\epsilon}.
\end{align}
Thus, the energy gap at weak magnetic field is given by
\begin{align}\label{eq:zigzaggap}
\Delta = \frac{\hbar v_F}{L}\frac{\sin 2\Theta}{2\sqrt{3} }\left(\frac{L}{l}\right)^2.
\end{align}
It grows linearly with magnetic field, and reaches the maximum in the dice model $\Theta=\frac{\pi}{4}$. 

\subsection{The \textbf{C}-\textbf{AB} boundary conditions.}

In the case of the C-AB boundary conditions we find that the spectra are asymmetrical with respect to the wave vector $k_0/2$. In
addition, we observe the formation of dispersionless surface solutions as $\Theta$ decreases. These solutions are bound to
the $k \simeq 0$ or $k \simeq k_0$ edges for the $K_{+}$ and $K_{-}$ valleys, similar to graphene [\onlinecite{Gusynin}] (see panel 
d in Fig.\ref{fig:zigzag_solution}). It is noticeable that they cease to be dispersionless at the opposite edges. This can be 
understood from the fact that the C-AB case of boundary conditions is similar to the zigzag boundary conditions in graphene. At small angles
$\Theta$ the lattice consists of hexagonal graphene-like part of A and C sites and weakly coupled triangle lattice of B sites. For C-AB ribbons the corresponding estimates of gap do not make sense because for different angles the gap is defined by different values of wave number 
$k$.

The spectrum at the large $L$ limit can be found in a similar manner to previous studied cases, and near boundary it is mainly determined by corresponding type of termination as in the half-plane system. Thus, the spectrum is given by \eqref{eq:ab_edge} for the edge electron states and \eqref{eq:ab_bulk} for the bulk states near the AB termination site as well as Eqs.\eqref{eq:c_edge} and \eqref{eq:c_bulk} near the C termination site, respectively.

\section{Armchair termination: a detailed derivations}
\label{appendix:armchair_add}
Here we give detailed derivation of results mentioned in Sec.\ref{sec:armchair} for ribbons with armchair termination.

Using the general form of solutions \eqref{eq:zigzag_general} for $\psi_C$ in both valleys with constants $A,\,B,\,A',\,B'$, we find the
following condition for the existence of nontrivial solutions:
\begin{align}
&\frac{4}{\pi}\cos\Delta KL -\nn &-\left(U'(\epsilon_1,\xi_1)V(\epsilon_2,\xi_1) + V'(\epsilon_2,\xi_1)U(\epsilon_1,\xi_1)\right)\left(U(\epsilon_2,\xi_2)V'(\epsilon_1,\xi_2) + V(\epsilon_1,\xi_2)U'(\epsilon_2,\xi_2)\right) +\nn
&+\left(V'(\epsilon_1,\xi_1)V'(\epsilon_2,\xi_1) + V'(\epsilon_1,\xi_1)V(\epsilon_2,\xi_1)\right)\left(U(\epsilon_1,\xi_2)U'(\epsilon_2,\xi_2) + U(\epsilon_2,\xi_2)U'(\epsilon_1,\xi_2)\right) +\\ 
&+\left(U(\epsilon_1,\xi_1)U'(\epsilon_2,\xi_1) + U'(\epsilon_1,\xi_1)U(\epsilon_2,\xi_1)\right)\left(V(\epsilon_2,\xi_2)V'(\epsilon_1,\xi_2) + V(\epsilon_2,\xi_2)V'(\epsilon_1,\xi_2)\right) - \nn
&-\left(V(\epsilon_1,\xi_1)U'(\epsilon_2,\xi_1) + V'(\epsilon_1,\xi_1)U(\epsilon_2,\xi_1)\right)\left(U(\epsilon_1,\xi_2)V'(\epsilon_2,\xi_2) + U(\epsilon_1,\xi_2)V(\epsilon_2,\xi_2)\right) = 0,\nonumber
\end{align}
where $\epsilon_1 = -\frac{\widetilde{\epsilon}^2}{4} + \frac{\cos2\Theta}{2}$,
$\epsilon_2 = -\frac{\widetilde{\epsilon}^2}{4} - \frac{\cos2\Theta}{2}$, $\xi_1 = \sqrt{2}k_yl$, $\xi_2 = \sqrt{2}(k_y l + L/l)$, and we used the
fact that the coefficient at $\cos\Delta KL$, being the Wronskian of parabolic cylinder functions, is constant, and equals $\sqrt{2/\pi}$ [\onlinecite{Abramowitz}].

Using the same method as for the zigzag BA-AB boundary conditions we find that for
$\cos \Delta KL = 1$ the applied magnetic field opens a gap which is described by the same equation as in the zigzag case
\eqref{eq:zigzaggap}. 

In the case $\cos\Delta KL = -1/2$ we apply the same method as for the zigzag C-C type of termination. After some 
algebraic simplifications we get
\begin{align}
\frac{4}{\pi} \cos\alpha - \frac{2 + 4\widetilde{\epsilon}^4}{\widetilde{\epsilon}^4\pi}\cos(\gamma_1 - \gamma_{-1})
+ \frac{2}{\widetilde{\epsilon}^4\pi}\left(\cos(\gamma_1 + \gamma_{-1}) - 2 \cos{4\Theta}\sin{\gamma_1}\sin{\gamma_{-1}} \right) = 0,
\end{align}
where $\gamma_{\mu} = \frac{k_0 l(12\mu\widetilde{\epsilon}^2 - \mu k_0^2 l^2 + 24\cos{2\Theta})}{12\sqrt{2\widetilde{\epsilon}^2 + 4\mu\cos{2\Theta}}}$.

The solution, which is accurate up to $B^2$ order, gives the gap magnitude
\begin{align}
\Delta= \frac{\hbar v_F}{L}\left[\frac{\pi}{3} + \frac{\sqrt{3}\pi^3 + 27(\sqrt{3}\pi - 9)\cos^2 2\Theta}{8\sqrt{3}\pi^4}\left(\frac{L}{l}\right)^4\right].
\end{align}
One should note that for $B=0$ this result coincides with the gap found in Ref.[\onlinecite{Oriekhov}]. Note that for small $\Theta$ the 
correction becomes negative. 

The approximate spectrum for edge and bulk electrons can be obtained in the same way as in zigzag case. 

\subsection*{Zero energy band}

For zero energy flat band $\widetilde{\epsilon}=0$, we seek solutions in the form
\begin{align}
&\psi_A = (\xi + \partial_\xi)\left[C_1 U\left(-a - \frac{\cos 2\Theta}{2}, \sqrt{2}\xi\right) + C_2V\left(- a- \frac{\cos 2\Theta}{2},\sqrt{2}\xi\right)\right],\\
&\psi_B = (\xi - \partial_\xi)\left[C_3 U\left(-a- \frac{\cos 2\Theta}{2}, \sqrt{2}\xi\right) + C_4V \left(- a- \frac{\cos 2\Theta}{2},\sqrt{2}\xi\right)\right],\\
&\psi_{A'} = (\xi - \partial_\xi)\left[C_5 U\left(-a+ \frac{\cos 2\Theta}{2}, \sqrt{2}\xi\right) + C_6V \left(- a+ \frac{\cos 2\Theta}{2},\sqrt{2}\xi\right)\right],\\
&\psi_{B'} = (\xi + \partial_\xi)\left[C_7 U\left(-a+ \frac{\cos 2\Theta}{2}, \sqrt{2}\xi\right) + C_8V \left(- a+ \frac{\cos 2\Theta}{2},\sqrt{2}\xi\right)\right],
\end{align}
where $a$ is an arbitrary real parameter.
Substituting them into the equations
\begin{align}
\cos\Theta(-\partial_\xi + \xi)\psi_A + \sin\Theta(-\partial_\xi - \xi)\psi_B = 0,\quad
\cos\Theta(\partial_\xi + \xi)\psi_{A'} + \sin\Theta(\partial_\xi - \xi)\psi_{B'} = 0.
\end{align}
Similar to the already studied zigzag case we find the following relations between constants $C_{1},\dots,C_{8}$:
\begin{align}
C_1 = C_3\tan\Theta \frac{a + \cos^2\Theta}{a - \sin^2\Theta},\quad C_2 = C_4\tan\Theta \frac{a + \cos^2\Theta}{a - \sin^2\Theta},\quad
C_5 = C_7\tan\Theta \frac{a-\cos^2\Theta}{a + \sin^2\Theta},\quad
C_6 = C_8\tan\Theta \frac{a-\cos^2\Theta}{a + \sin^2\Theta}.
\end{align}
The particular cases where $a=\cos^2\Theta$ or $a=\sin^2\Theta$ result in trivial solutions for wave functions.

Substituting the obtained nontrivial solutions into the boundary conditions \eqref{eq:armchair_conditions} and setting the determinant of the
system to zero, we find the characteristic equation. Its solutions are shown in Fig.\ref{fig:armchair_zeroband}.

\section{Asymptotic expansions of parabolic cylinder functions}
\label{sec:UV_asymptotic}
Here we present main asymptotic formulas for parabolic cylinder functions $U(a,x)$ and $V(a,x)$, which where used throughout the paper. 

Firstly one should note the relation between $U$ and $V$ functions with more convendional $D_{\nu}(x)$ function:
\begin{align}
	&U(a, z)=D_{-a-1 / 2}(z), \quad V(a, z)=\frac{\Gamma(a+1 / 2)}{\pi}\left[\sin (\pi a) D_{-a-1 / 2}(z)+D_{-a-1 / 2}(-z)\right].
\end{align}
The relations between $U(a,x)$, $V(a,x)$ with positive and negative arguments are the following:
\begin{align}\label{eq:U_minus_x}
	&U(a,-x)=\frac{\pi  V(a,x)}{\Gamma \left(a+\frac{1}{2}\right)}-\sin (\pi  a)
	U(a,x),\\
	\label{eq:V_minus_x}
	&V(a,-x)=\frac{\cos ^2(\pi  a) \Gamma \left(a+\frac{1}{2}\right) U(a,x)}{\pi }+\sin
	(\pi  a) V(a,x).
\end{align}
The Taylor series at fixed $a$ and small arguments are given by 
\begin{align}\label{eq:Taylor_expansion1}
&	U(a, x) = \frac{2^{-\frac{1}{4} - \frac{a}{2}}\sqrt{\pi}}{\Gamma\left(\frac{1}{2} + \frac{1}{2}\left(\frac{1}{2} + a\right)\right)}\left(1 + \frac{1}{2}a x^2 +\frac{2a^2 + 1}{48}x^4\right) - \frac{2^{\frac{1}{4} - \frac{a}{2}}\sqrt{\pi}}{\Gamma\left(\frac{1}{2}\left(\frac{1}{2} + a\right)\right)} \left(x + \frac{a}{6}x^3\right) + O\left(x^5\right),\\
\label{eq:Taylor_expansion2}
&	V(a, x) = \frac{2^{-\frac{1}{4} - \frac{a}{2}}\Gamma\left(a + \frac{1}{2}\right)(\sin{a\pi} + 1)}{\sqrt{\pi}\Gamma\left(\frac{1}{2} + \frac{1}{2}\left(\frac{1}{2} + a\right)\right)}\left(1 + \frac{1}{2}a x^2 +\frac{2a^2 + 1}{48}x^4\right) - \frac{2^{\frac{1}{4} - \frac{a}{2}}\Gamma\left(a + \frac{1}{2}\right)(\sin{a\pi} - 1)}{\sqrt{\pi}\Gamma\left(\frac{1}{2}\left(\frac{1}{2} + a\right)\right)} \left(x + \frac{a}{6}x^3\right) + O\left(x^5\right).	
\end{align}
The asymptotic formulas at fixed $a$ and $x\to\infty$ are different depending on argument sign: 
\begin{align}\label{eq:asympxU}
&U(a,x)=e^{-\frac{x^2}{4}} x^{-a-\frac{1}{2}},\quad U(a,-x)=\frac{\sqrt{2 \pi } e^{\frac{x^2}{4}} x^{a-\frac{1}{2}}}{\Gamma \left(a+\frac{1}{2}\right)},\\
\label{eq:asympxV}
&V(a,x)=\sqrt{\frac{2}{\pi}}e^{\frac{x^2}{4}} x^{a-\frac{1}{2}}, \quad V(a,-x)=\sqrt{\frac{2}{\pi}}e^{\frac{x^2}{4}} x^{a-\frac{1}{2}} \sin (\pi  a).
\end{align}
The derivatives can be estimated as follows:
\begin{align}\label{eq:asymp_derivU}
&U'(a, x) = -\frac{1}{2} e^{-\frac{x^2}{4}} x^{-a+\frac{1}{2}},\quad
U'(a, -x) = -\sqrt{\frac{\pi}{2}}\frac{ e^{\frac{x^2}{4}} x^{a+\frac{1}{2}}}{\Gamma \left(a+\frac{1}{2}\right)},\\
\label{eq:asymp_derivV}
&V'(a, x) = \frac{1}{\sqrt{2 \pi }} e^{\frac{x^2}{4}} x^{a+\frac{1}{2}},\quad
V'(a, -x)= -\sqrt{\frac{1}{2 \pi }} e^{\frac{x^2}{4}} x^{a+\frac{1}{2}} \sin (\pi  a). 
\end{align}
Both functions are real if $a$ and $x$ parameters are real [\onlinecite{Bateman}]. Using this fact, we obtain the following asymptotic formulas in the case $|a|\gg x^2$, for $a<0$:
\begin{align}\label{eq:largea1}
&U(a, x) = \frac{\Gamma\left(\frac{1}{4} - \frac{a}{2}\right)}{2^{\frac{a}{2} + \frac{1}{4}}\sqrt{\pi}}\left(1 - \frac{x^2}{16a}\right)\cos\left(\pi\left(\frac{a}{2} + \frac{1}{4}\right) + \sqrt{-a}x - \frac{x^3}{24\sqrt{-a}}\right), \\
&V(a, x) = \frac{\Gamma\left(\frac{1}{4} - \frac{a}{2}\right)}{\Gamma\left(\frac{1}{2}-a\right)2^{\frac{a}{2} + \frac{1}{4}}\sqrt{\pi}}\left(1 - \frac{x^2}{16a}\right)\sin\left(\pi\left(\frac{a}{2} + \frac{1}{4}\right) + \sqrt{-a}x - \frac{x^3}{24\sqrt{-a}}\right), \\
&U'(a, x) = -\frac{\Gamma\left(\frac{3}{4} - \frac{a}{2}\right)}{2^{\frac{a}{2} - \frac{1}{4}}\sqrt{\pi}}\left(1 + \frac{x^2}{16a}\right)\cos\left(\pi\left(\frac{a}{2} + \frac{1}{4}\right) + \sqrt{-a}x - \frac{x^3}{24\sqrt{-a}}\right), \\
\label{eq:largea4}
&V'(a, x) = -\frac{\Gamma\left(\frac{3}{4} - \frac{a}{2}\right)}{\Gamma\left(\frac{1}{2}-a\right)2^{\frac{a}{2} - \frac{1}{4}}\sqrt{\pi}}\left(1 + \frac{x^2}{16a}\right)\sin\left(\pi\left(\frac{a}{2} + \frac{1}{4}\right) + \sqrt{-a}x - \frac{x^3}{24\sqrt{-a}}\right),
\end{align}
which can be derived from relations 19.11.1 and 19.11.2 from the book [\onlinecite{Abramowitz}].

\end{document}